\documentclass[letterpaper,9pt, twocolumn]{extarticle}    
\usepackage[letterpaper, dvips=true, scale=0.862]{geometry}
\usepackage[utf8]{inputenc}
\usepackage[mcite]{achemso}
\usepackage{amsmath}
\usepackage[dvips]{graphicx}

\usepackage[roman]{sublabel}

\makeatletter
\renewcommand\section{\@startsection {section}{1}{\z@}%
       {-1.5ex \@plus -.5ex \@minus -.8ex}%
       {1.5ex \@plus.2ex \@minus .2ex}%
       {\raggedright\normalfont\large\bfseries\sffamily}}
    \renewcommand\subsection{\@startsection{subsection}{2}{\z@}%
       {-1ex\@plus -.4ex \@minus -.4ex}%
       {1ex \@plus .2ex \@minus .2ex}%
       {\raggedright\normalfont\bfseries\sffamily}}
    \renewcommand\subsubsection{\@startsection{subsubsection}{3}{\z@}%
      {2ex \@plus1ex \@minus.3ex}%
      {-1em}%
      {\normalfont\normalsize\bfseries\sffamily}}
\makeatother
\newcommand{\degree}{\ensuremath{\,^{\circ}\mathrm{C}}}

\title{Homochirality and the need of energy}
\author{\emph{Rapha\"el Plasson$^1$\thanks{\texttt{rplasson@nordita.org}},
   Axel Brandenburg$^{1,2}$}\\
  $^1$Nordita, Stockholm, Sweden, $^2$Department of Astronomy,
  Stockholm University, Sweden.}

\date{}

\begin{document}

\maketitle

\begin{abstract}
  The mechanisms for explaining how a stable asymmetric chemical
  system can be formed from a symmetric chemical system, in the absence of
  any asymmetric influence other than statistical fluctuations, have
  been developed during the last decades, focusing on the non-linear
  kinetic aspects. Besides the absolute necessity of
  self-amplification processes, the importance of energetic aspects is
  often underestimated. Going down to the most fundamental aspects,
  the distinction between a single object --- that can be
  intrinsically asymmetric --- and a collection of objects --- whose
  racemic state is the more stable one --- must be emphasized.  A
  system of strongly interacting objects can be described as one
  single object retaining its individuality and a single asymmetry;
  weakly or non-interacting objects keep their own individuality, and
  are prone to racemize towards the equilibrium state. In the presence of
  energy fluxes, systems can be maintained in an asymmetric
  non-equilibrium steady-state. Such dynamical systems
  can retain their asymmetry for times longer than their racemization
  time.

  \paragraph{Keywords:} Emergence of Homochirality $\cdot$ Energy $\cdot$ Entropy $\cdot$
    Non-Equilibrium Thermodynamics $\cdot$ System Chemistry $\cdot$ Symmetry
  Breaking
\end{abstract}

\section*{Introduction}
\label{sec:introduction}

Understanding how asymmetric chemical systems can be formed
spontaneously is one key element for the description of the onset of
life. The emergence of homochirality has been developed as a kinetic
phenomenon, describing chemical competition in the presence of
autocatalytic phenomena \cite{Frank-53,Kondepudi.Nelson-83}. In this
context, asymmetric states can be described as originating from
statistical fluctuations, amplified and maintained actively in
non-equilibrium steady states thanks to high-order autocatalytic
reactions \cite{Plasson.Kondepudi.ea-07}.

In recent years the interest in energetic aspects of homochirality has
become apparent
\cite{Kondepudi.Kapcha-08,Blackmond.Matar-08,Plasson.Bersini-09}. Assuming
the presence of efficient autocatalytic processes, we will mainly
focus here on the energetic aspects that must be taken into account in
a far-from-equilibrium context. By focusing on simple examples that
are partly motivated by recent experiments of different groups, we
address the question where the energy responsible for the symmetry
breaking behaviors originates from.

We shall first discuss the significance of symmetry versus asymmetry
of chemical systems, emphasizing the underlying apparent paradox of
homochiral phase stability that is the source of many
misunderstandings. Developing the dynamic aspect of asymmetric
systems, a detailed picture about the energetic requirements for their
generation is given. In particular, the necessity of energy for
creating microscopic and macroscopic asymmetry, and for maintaining
this asymmetry in a stable steady state are highlighted. Concrete
examples from the literature will be developed to illustrate these
energy needs.

\section*{Symmetry of chemical systems}
\label{sec:chem-sym}

We shall assume that the fundamental laws of physics are
symmetric\footnote{This is of course not strictly true, because of the
  existence of the violation of symmetry by weak
  force\cite{Mason.Tranter-85,Chandrasekhar-08,Pagni-09}. However, it
  is widely accepted in the community that the influence of this
  asymmetry can be neglected in the presence of autocatalytic
  mechanisms, that only require initial statistical fluctuations to
  bootstrap the
  process\cite{Avalos.Babiano.ea-00,Bonner-00,Lahav.Weissbuch.ea-06,Lente-07}.},
the purpose of this article being to describe how asymmetric states
can emerge from symmetric laws. Following Curie's Principle
\cite{Curie-94}\footnote{``Lorsque certaines causes produisent
  certains effets, les éléments de symétrie des causes doivent se
  retrouver dans les effets produits. Lorsque certains effets révèlent
  une certaine dissymétrie, cette dissimétrie doit se retrouver dans
  les causes qui lui ont donné naissance. La réciproque de ces deux
  propositions n'est pas vraie, au moins pratiquement, c'est-à-dire
  que les effets produits peuvent être plus symétriques que les
  causes.'' (When some causes produce some effects, the elements of
  symmetry of the causes must be found in the produced effects. When
  some effects possess some asymmetry, this asymmetry must exist in
  the causes that produced the effects. The reciprocal of these two
  propositions is not true, at least in practice, that is the produced
  effects can be more symmetric than the causes.)}, this may seem at
the first glance impossible, but because of the discrete nature of
matter, statistical fluctuations are an unavoidable source of
asymmetry, apt to be amplified by appropriate mechanisms
\cite{Siegel-98,Plasson.Kondepudi.ea-07}.

 And indeed, it is easy to realize that asymmetry can be naturally
 obtained as long as a chemical compound is sufficiently complex. It
 only requires four different groups to be attached to one atom of
 carbon to obtain a chiral molecule. It is sufficient to build
 \emph{one} sufficiently complex molecule to generate
 \emph{microscopic} asymmetry. However, this is not sufficient to
 build macroscopically ordered asymmetry. A sufficiently large
 collection of objects will lose its ordered asymmetry: Curie's
 Principle suggests that, on average, the same quantity of both
 configurations will be created. Here appears an apparent paradox: one
 object is asymmetric, one collection of object is symmetric. How can
 this problem be solved?

\subsection*{Symmetry of an object}
\label{sec:sym-obj}

An object is asymmetric if it is not similar to its image by the
corresponding symmetry operation. In the present case, we are talking
of chiral objects, that are different from their mirror image.
This can be applied to amino acids, nucleotides, crystals, and
other objects, but to keep it general we shall now talk about
``$+$'' and ``$-$'' compounds (that
can represent R and  S compounds, P and M helices, etc.).

\begin{figure}[htb]
  \centering
    \includegraphics[width=4cm]{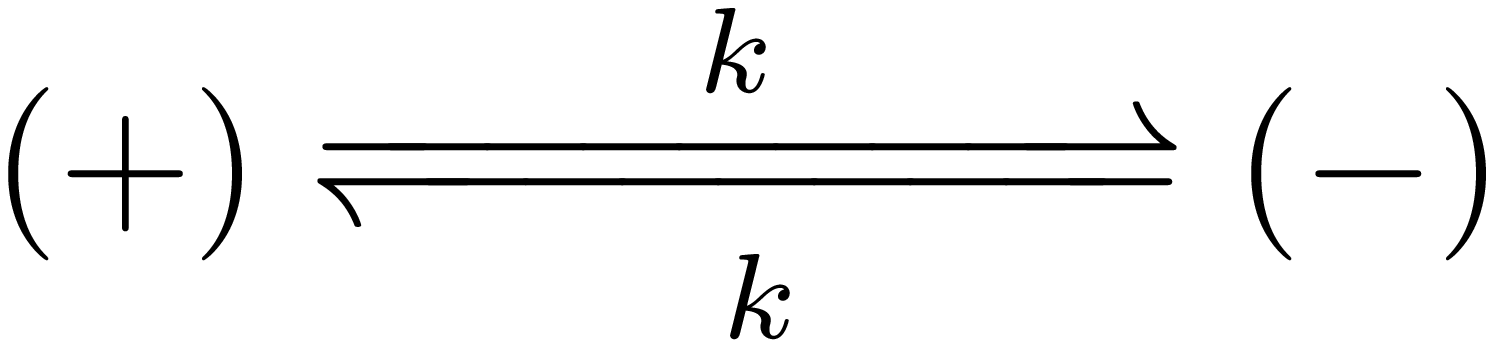}\\
    \includegraphics[width=4cm]{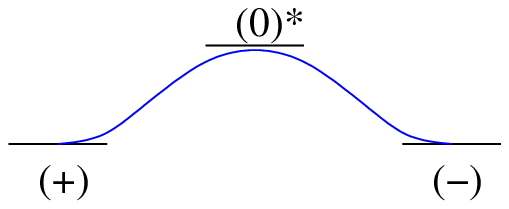}%
  \caption{Temporal conversion of a chiral molecule. It stays half of its
    life time in each of the two configurations on a timescale larger that the
    reaction half time, passing by a symmetric intermediate $(0)^*$.}
  \label{fig:time_chiral}
\end{figure}

Because of the symmetry of physical laws, a chiral molecule is exactly
as stable as its enantiomer (its mirror image molecule). It will
interconvert to its mirror image on a given timescale (that may be
very small or very large). The kinetic rate (or inversion probability)
is the same in both directions for symmetry reason. In practice, the
asymmetry of an object persists only on a timescale smaller than its
interconversion time. At larger timescale, they spend half of the time
in one configuration, and half of the time in the other configuration
(see Fig.~\ref{fig:time_chiral}). They are symmetric on average,
despite being pointwise asymmetric on a given timescale. This is
typically the case of tri-substituted amines, or of the asymmetric
conformation of symmetric molecules, that are theoretically asymmetric
but that do not possess any chiral properties at ambient
temperature. Strictly speaking, a chiral object should thus be
considered as in a frozen state, whose asymmetry exists only
temporarily. However, this frozen state is purely academic in the
cases where the rate of inversion can be neglected.  This is
especially the case for macroscopic objects, like crystals, whose
spontaneous inversion is not possible\footnote{Some precautions must
  be taken there. For example, if a crystal of L-amino acid cannot
  invert into the mirror crystal of D-amino acid, amino acids can
  racemize in solid form (non-crystalline), even if this process is
  much slower than in solution. For example, while the half life of
  aspartic acid in $100\degree$ water is 30 days in free form and 1
  to 3 days in proteins, it will still racemize on timescales of
  $10^4-10^5$ years in solid form, like in fossils or carbonate
  sediments \cite{Bada-85}}.

\subsection*{Symmetry of a collection of objects}
\label{sec:sym-coll}

Real systems, and especially chemical systems, will generally not be
composed of one unique object, but of a large number of them. Let us
take a system composed of $n$ chiral objects, having $m$ of ``$+$''
type and $(n-m)$ of ``$-$'' type. There is a large number of different
ways to build such a system. The number of equivalent combinations is:
\begin{equation}
  \label{eq:combin}
  c_m= \frac{n!}{(n-m)!m!}.
\end{equation}
There are $c_m$ equivalent systems of $n$ objects with $m$ of ``$+$''
type, each having the same probability $p_m=1/c_m$, and their entropy
is \cite[p 59, Eq 2.46]{Bellac.Mortessagne.ea-04}:
\begin{eqnarray}
  \label{eq:entropy}
  S_m &=& -\sum_{k=1}^{c_m} p_m \ln p_m\\
  &=& \ln c_m.
\end{eqnarray}
This value is zero for homochiral systems (i.e. for $m=0$ or $m=n$),
and maximum for racemic systems (i.e. for $m=n/2$). This implies an
entropic stabilization of the racemic state, equal to:
\begin{eqnarray}
  \label{eq:entro_stab}
  \Delta S&=&S_{n/2}-S_0\\
  &=&\ln\frac{n!}{
    \left(
      \frac{n}{2} !
    \right)^2}.
\end{eqnarray}
For very large systems, this can be evaluated by using the Stirling
approximation, $\ln(n!)\approx n\ln n -n$:
\begin{eqnarray}
  \label{eq:entro-stab-large}
  \Delta S&=&\ln(n!)-2\ln
  \left(
    \frac{n}{2}!
  \right)\\
  &\approx&n\ln n -n -2
  \left(
    \frac{n}{2} \ln \frac{n}{2} - \frac{n}{2}
  \right)\\
  &=&n\ln2.
\end{eqnarray}
The corresponding macroscopic molar entropy of the racemic state is:
\begin{eqnarray}
  \label{eq:entro-stab-macro}
  S&=&k_b\mathcal{N}_a\ln2\\
  &=&R\ln 2=5.76~\mathrm{J}\,\mathrm{K}^{-1}\,\mathrm{mol}^{-1},
\end{eqnarray}
where $k_b$ is the Boltzmann constant, $\mathcal{N}_a$ the Avogadro's
number, and $R$ the ideal gas constant. This value corresponds to the
entropy of mixing of half a mole of compounds ``$+$'' and of half a
mole of compounds ``$-$''. \footnote{It may seem surprising that a
  thermodynamic stabilization is obtained despite the absence of any
  interaction between the compounds, and that the value is independent
  of the nature of the compounds. This is referred in the literature
  as the ``Gibbs Paradox'', and many interpretations have been
  developed
  \cite{Jaynes-92,Lin-96,Ben-Naim-07,Perezmadrid-04,Lin-08,Swendsen-08}.}

\begin{figure}[htb]\centering  
  \includegraphics[width=9cm]{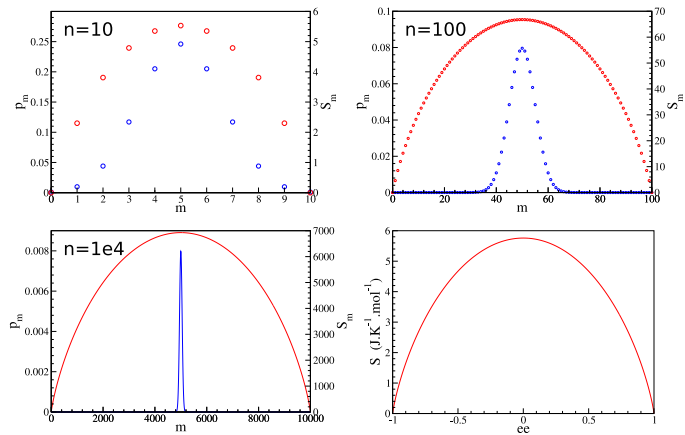}
  \caption{Entropic stabilization (red) and statistical
    distribution (blue) for a non-interacting ensemble of $n=10$, $100$
    and $10000$ chiral objects containing $m$ configurations of type $+$ and
     $(n-m)$ configurations of type $-$, and entropy of mixing for a macroscopic
    system of chiral objects as a function of the enantiomeric
    excess. The statistical distribution of a macroscopic system is too narrow to
    be represented.}
\label{fig:entro-stab}
\end{figure}

In total, there are $\sum_{m=1}^nc_m=2^n$ systems of $n$
compounds. All these systems are energetically equivalent because of
the symmetry between $+$ and $-$ compounds. Among all these
systems, for each $m$, there are $c_m$ systems that are totally
equivalent (i.e.\ that cannot be distinguished from each other). The
proportion of systems having $m$ of type $+$ is $c_m/2^n$: that represents a
Gaussian distribution of the systems of $n$ compounds, centered on
$m=n/2$, with a standard deviation $\sigma=\sqrt{n}/2$. The
racemic state is the more stable state, as it is entropically
stabilized, and the statistical fluctuations correspond to a standard
deviation for the enantiomeric excess of $\sigma_{ee}=1/\sqrt{n}$
 \cite{Hill-62,Siegel-98,Faglioni.Passalacqua.ea-05}.

\subsection*{The paradox of homochiral phase stability}
\label{sec:parad-homoch-phase}

The entropic stabilization relies on the fact that there are no
enthalpic effects, i.e.\ that there are no interactions between the
objects. But even if both configurations have the same thermodynamic
properties for symmetry reason, the system is more complex when
interactions are possible. While retaining the global symmetry of the
system, a compound $+$ will interact differently with another $+$
or with a $-$ compound. The symmetry of the laws only implies that the
interactions involving $++$ and $--$ are identical (and likewise for
the interactions involving $+-$ and $-+$) but interactions involving
$++$ and $+-$ may differ.

If heterochiral interactions are favored, then the system evolves to a
symmetric phase, each favored $+-$ association losing its
asymmetry (e.g.\ in a racemate crystal). If homochiral interactions are
favored, the system evolves to an asymmetric phase (e.g.\ a
conglomerate crystal). However, the situation did not evolve much:
either one unique homochiral phase is obtained (e.g.\ a unique racemate
monocrystal), or several independent phases (e.g.\ a racemic mixture of
racemate monocrystals). The individual objects are interacting, and thus
form new larger independent objects to which the entropic
stabilization still applies, so the thermodynamic equilibrium is still
a racemic mixture of independent asymmetric objects\footnote{At first
  glance, this may seem to contradict the application of the Gibbs
  phase rule to these systems, that describes the homochiral phase as
  the stable one
  \cite{Crusats-06,Blackmond-07*d,Crusats-07,Blackmond-07*b}. However,
  the Gibbs phase rule comes down to determining how many degrees of
  freedom are available in the system, given the number of parameters
  and available relations between them. In the context of crystals,
  this calculation does not take into account the number of crystals
  of the same configuration and their size distribution. This
  presupposes that, as a working hypothesis, these parameters are not
  relevant for describing the properties of the system. In that
  context, a system with one large crystal is considered equivalent to
  a system with a large number of small crystals.  The Gibbs phase
  rule can only be applied to describe the stability of one unique
  phase, unless the parameters describing the distribution of crystals
  are introduced. This explains our assumption of independent
  asymmetric objects.}. An intermediate case exists in the case of
weak stereoselectivity, as, for example, in the system of
crystallization of 1,1'-binaphtyl \cite{Asakura.Soga.ea-02}. Stirred
crystallization of these compounds generally leads to very noisy
distributions, and a single crystal grown from a single seed is not
fully homochiral. In this system, the homochiral interactions are only
weakly favored, causing the possibility to have inclusions of crystals
of the opposite configuration during the crystal growth
\cite{Asakura.Nagasaka.ea-04,Plasson.Kondepudi.ea-06,Asakura.Plasson.ea-06}.
This is characterized by the association of several objects (crystal
inclusions) weakly interacting between each other within one unique
object (the full-grown crystal) that is less asymmetric than
expected. Similar systems generating partially homochiral systems can
be found, typically in the case of polymers \cite{Hitz.Luisi-03}.

When the interactions are only temporary, the symmetric compounds
conserve their individuality, and they will stay racemic \emph{at
  equilibrium}. Due to the microreversibility of reactions, direct
racemization is not necessary: if a pathway exists between the two
enantiomers, it will be equivalent to a racemization
reaction. Typically, in the system of crystallization of amino acids
of Noorduin \emph{et al}\cite{Noorduin.Meekes.ea-08,Noorduin.Meekes.ea-08*b} a homochiral
system of objects is created (crystals), but the chiral compounds keep
their individuality in solution, which thus remains racemic.

In practice, the interactions will create more or less large
``homochiral objects'' (like molecules, homochiral blocks within
polymers, entire polymers, crystals, etc.) that have a cohesion.
Thus, each object is enthalpically stabilized and can thereby overcome
the entropic influence. On a large enough scale, the collection of
objects will anyway still be globally racemic (i.e.\ only local
homochirality can be obtained).  This does not mean that it is
impossible to obtain a stable non-racemic mixture of individual
objects, but only that this is impossible in the \emph{equilibrium}
state: there is a necessity for maintaining the system in
non-equilibrium, that is, there is a necessity of energetic exchange.

\section*{Systems evolution and stability}
\label{sec:evol-stab}

The previous description was essentially static, describing what is
the more stable state of the system. Let us now develop the dynamic
description. We start from a symmetric system that may be either
composed of only symmetric objects or of a racemic mixture of
asymmetric objects. The purpose here is to understand how it can
evolve towards an asymmetric stable steady state.

\subsection*{Creating an asymmetric state}
\label{sec:create}

The first step is to create the asymmetry. This can be a system that
creates asymmetric objects either from symmetric objects or by
the transformation of asymmetric objects into other asymmetric
objects. Alternatively, it can be a system with interconversion between
different enantiomers of asymmetric objects.

The first case implies the presence of a source of energy that can be
internal (the starting compounds are less stable, allowing their
spontaneous transformation) or external (the transformations are
coupled to energy exchange allowing the reactions to be performed in
the appropriate direction\footnote{How chemical energy can be
  transferred is explained briefly in the Appendix. The general idea
  is to couple an energetically unfavored reaction to an energetically
  favored reaction. This results in a transfer of chemical energy,
  allowing the system to perform a reaction that cannot be
  spontaneously performed.}).  The second case implies the existence
of a pathway allowing the interconversion between asymmetric objects,
that is, a racemization process: such a system will reach a racemic
state at equilibrium, and must thus be maintained away from this
equilibrium, which will require energy.

\begin{figure}[htb]
 \centering
  \includegraphics[width=3cm]{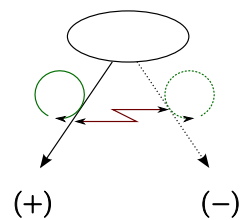}%
\caption{Generation of an asymmetric state, starting from a symmetric
state (either constituting of asymmetric compounds or of a racemic mixture
of symmetric compounds). The generation of the asymmetric compounds
must be linked to autocatalytic processes (half-open arrowed rings) and
to mutual inhibition (the double-sided zig-zag arrow)}
  \label{fig:gen}
\end{figure}

In any case, an energy consuming process must first be established in
order to create an asymmetric state. As the system is initially
totally symmetric, a bifurcation towards one or the other
configuration is required. If the process leads to an unique
asymmetric object (i.e. if all the compounds get aggregated), then the
final state can only be homochiral. In other cases, self-amplification
processes are necessary. There must then be an autocatalytic formation
of the asymmetric compounds (in order to enhance the difference
between both configurations) and a mutual inhibition process (in order
to destroy compounds of the ``bad'' configuration). This process is
represented in Fig.~\ref{fig:gen}.  This is the essence of the Frank's
model \cite{Frank-53,Kondepudi.Nelson-83}. These processes may be
indirectly generated within the chemical network, rather than being
direct reactions
\cite{Plasson.Bersini.ea-04,Plasson.Kondepudi.ea-07,Brandenburg-07}.

\subsection*{Maintaining an asymmetric state}
\label{sec:maintain}

In the presence of a racemization process, the system will tend to destroy
the previously built asymmetry, and go back to the equilibrium
state. However, if the process that was used to build the asymmetry in
the first place can be continuously maintained, then it may be able to
oppose the spontaneous racemization process, resulting in the
maintenance of a non-equilibrium asymmetric steady state
\cite{Plasson-08,Plasson.Bersini-09}.

\begin{figure}[htb]
   \centering
    \includegraphics[width=8cm]{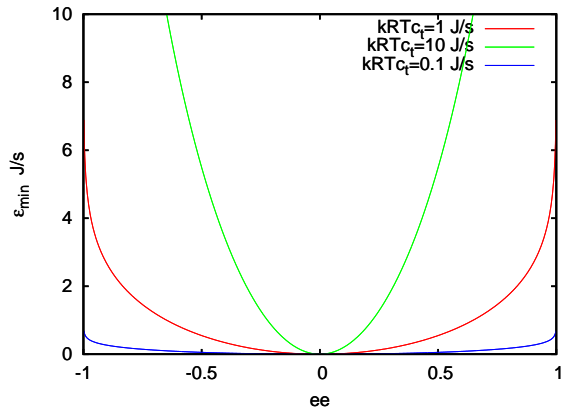}%
  \caption{Profile of the minimal energy required for maintaining a
    given $ee$ for different values of the product $kRTc_t$.}
  \label{fig:min_energ}
\end{figure}

This process cannot be costless in term of energy. If an asymmetric
state, characterized by an enantiomeric excess $ee$, is maintained in
a steady state, it means that different concentrations $c_+$ and $c_-$
of $+$ and $-$ will be maintained. As a consequence, the
racemization reaction (with an identical kinetic constant $k$ in both
directions, because of the microreversibility), is not anymore
in detailed balance, as $kc_+ \neq kc_-$. This generates a continuous
reaction flux, maintained by the incoming energy, that will
continuously lead to an entropy production $\sigma$. The incoming
energy flux $\varepsilon$ must at least compensate $T\sigma$, that is: 
\begin{eqnarray}
  \sigma&=&R(kc_+-kc_-)\ln\frac{kc_+}{kc_-} \label{eq:ent_prod}\\
  \Rightarrow \qquad \varepsilon& \geq & kRTc_t \cdot ee \ln\frac{1+ee}{1-ee} \label{eq:energ_cons}
\end{eqnarray}
with $c_t=c_++c_-$ and $ee=(c_+ - c_-)/c_t$.  More energy will be required for systems with
fast racemization, high maintained enantiomeric excess, and high
concentrations in compounds. Inversely, no energy is required for
maintaining racemic systems ($ee=0$) or frozen systems ($k=0$). It can
be noted that for maintaining strictly homochiral systems ($ee=\pm
1$), 
an infinite amount of energy is required, whatever the value of
$k$.  This is compensated by the fact that the mathematical function
describing $\varepsilon$ is very steep when approaching $\pm 1$ (See
Fig.~\ref{fig:min_energ}).

Such dynamic systems are fundamental to obtaining stable systems of
non-racemic compounds under racemizing conditions, as this is typically
the case for amino acids and peptides in aqueous solution. In this
domain, biosystems are especially efficient: while racemization of
amino acid derivatives in living matter can occur on the
timescale of its life span \cite{Masters.Bada.ea-77}, the process
of life managed to maintain a very high enantiomeric excess in organic
compounds during billion of years.

\subsection*{Propagating the asymmetry in space}
\label{sec:spat-extend-models}

The chemical processes in realistic settings take place in a spatially
extended domains (chemical reactors, in oceans, near shore lines, in
the atmosphere, in the porous interface to the Earth crust, etc).  In
at least some of those cases it is possible to treat the chemistry by
adding spatial diffusion and/or advection terms.  The system is thus
described as a collection of local systems, able on their own to
consume energy for generating and maintaining an asymmetric state as
described above, and that can additionally interact with neighboring
local systems and influence them, or be influenced by them. The
relevant equations take then the form of partial differential
equations of the form
\begin{equation}
{\partial X_i\over\partial t}+\vec{u}\cdot\vec{\nabla} X_i
-\kappa_i \nabla^2 X_i = \mbox{reaction terms},
\end{equation}
where the concentrations $X_i$ are not only functions of time, but
also of spatial coordinates $\vec{r}$.  Here, $u(\vec{r},t)$ is the advective
velocity and $\kappa_i$ is the molecular diffusivity.

In systems with only reaction-diffusion equations, the typical
solutions tend to be of the form of propagating fronts.  This behavior
seems to be quite general and independent of the detailed chemistry,
and have now been obtained by various groups
\cite{Saito.Hyuga-04*b,Brandenburg-04,Multamaki.Brandenburg-05,Hochberg-06,Gleiser-07,Gleiser.Walker-09}.

Initially, the diffusion does not have much time to operate: in the
initially racemic mixture some local systems bifurcate to one
configuration, and some other bifurcate to the opposite one, while the
total system remains globally racemic.  These different regions are
initially well separated and will then spread by simple diffusion into
the still racemic regions.  This process may already be quite slow,
but what happens next is even slower, because once two regions of
opposite chirality come into contact, the propagation of the front
slows down.  This leads to a system with homochiral regions, separated
by thin racemic walls.

The regions in contact will compete with each other. Their interface
propagates in the direction of the curvature, because the inner front
is shorter than the outer one; linear interfaces are stables. The
propagation speed is constant in time, and therefore the globally
averaged enantiomeric excess grows linearly in time, and is
proportional to the number of the still topologically separated
regions in space. Eventually, different regions in space merge, so the
number of topologically separated regions in space decreases with
time, and the growth of the global e.e.\ slows down. Eventually, if the
system is of finite size, one single configuration may be established
everywhere, but several space regions of opposite configuration can
coexist, separated by stable racemic walls.

Here again, if each local system can be considered as single objects
when isolated, they can be fused with other objects (the neighboring
systems) to generate larger objects (the homochiral zones).
In the final stable state there is either one completely interconnected
homochiral zone (i.e.\ one single object), or there is a coexistence of
several zones (i.e.\ multiple objects) that will be globally racemic
for sufficiently large systems. This propagation of the asymmetry can
be enhanced by the advection phenomenon: adding mechanical motions able to
stir the system will perturb the interfaces and increase the
transport speed of compounds inside the system, leading thus much faster to
one unique configuration in all space locations
\cite{Brandenburg-04,Multamaki.Brandenburg-05}. The effects of noise,
which corresponds to a stochastic loss of chirality in random
locations, can however offset this effect either partially or completely
\cite{Gleiser-07,Gleiser.Walker-09}.

\subsection*{The need of energy}
\label{sec:need_energy}

\begin{figure}[htb]
  \centering
  \parbox[t]{3cm}{\centering%
    \includegraphics[width=2.4cm]{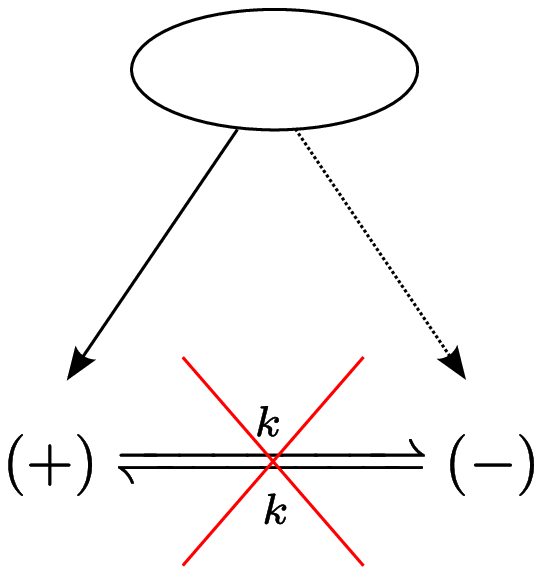}\\
    A. Frozen System%
  }%
  \parbox[t]{3cm}{\centering%
    \includegraphics[width=2.4cm]{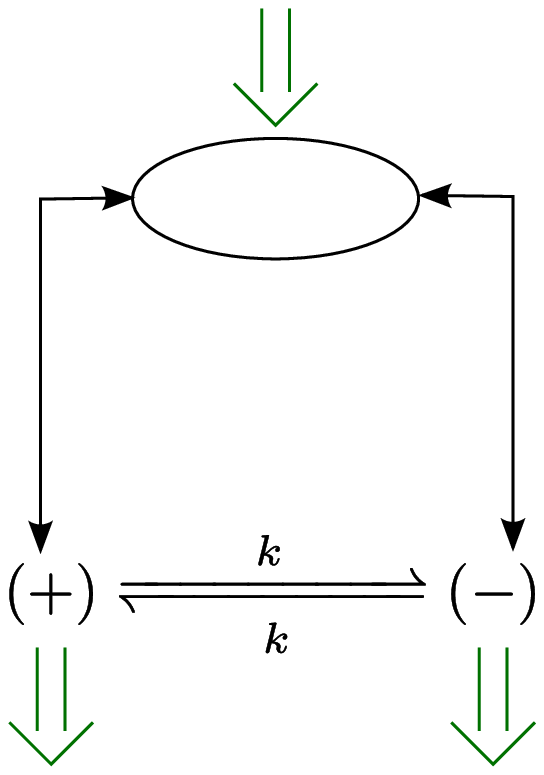}\\
    B. Open-flow System%
  }%
  \parbox[t]{3cm}{\centering%
    \includegraphics[width=2.4cm]{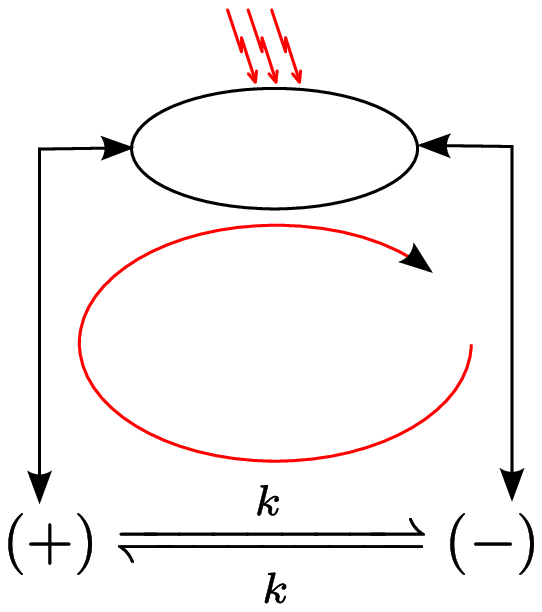}\\
    C. Recycled System%
  }
  \caption{Three frameworks for obtaining a stable non-racemic
    state. A: an isolated system that can evolve from a symmetric
    state to an asymmetric state may be stable as long as the
    racemization process can be neglected. B: An open flow system can
    maintain an asymmetric steady state, refueling continuously in
    freshly synthesized asymmetric compounds, as long as a complete
    ``flush'' of the system is avoided
     \cite{Cruz.Parmananda.ea-08}. C: A non-isolated recycled system
    can maintain an activated deracemization process, opposed to the
    spontaneous racemization process  \cite{Plasson-08}.}
  \label{fig:stabl_non_rac}
\end{figure}

Fig.\ref{fig:stabl_non_rac} gives the main frameworks for obtaining a
stable non-racemic state. The fundamental need of energy can
originate from an initial excess of chemical energy
(Fig.~\ref{fig:stabl_non_rac}A), from an exchange of matter with the
surroundings (Fig.~\ref{fig:stabl_non_rac}B), or from an exchange of
energy with surroundings (Fig.~\ref{fig:stabl_non_rac}C).

The first category corresponds to the frozen systems. They are static
systems that are initially placed in a non-equilibrium state,
possessing an excess of free internal energy, and that evolve
spontaneously towards a more stable state. No more processes happen in
this final state, as all energy has been consumed, and no further
exchanges occur. This system will be prone to processes of
racemization and is bound to racemize in the long term. The asymmetric
state is formed under kinetic control, but the symmetric state will be
slowly reached under thermodynamic control
\cite{IUPAC.McNaught.Wilkinson-97}, as the real equilibrium state .

The second and third categories are the dynamic systems: the final
steady state is non-isolated, continuously exchanging energy or matter
with its surroundings. This minimum of required energy is actually very
low. For example, taking the racemization rates of aspartic acid from
Bada\cite{Bada-72}, it corresponds to a minimum consumption of energy of
$\varepsilon_{min}=1.3\cdot10^{-7}~\mathrm{J}\,\mathrm{s}^{-1}$ per mole of amino acid
maintained in an enantiomeric excess of $99\%$ at $25\degree$, and of
$1.6\cdot 10^{-3}~\mathrm{J}\,\mathrm{s}^{-1}$ at $100\degree$.  In acidic or basic
conditions, kinetic rates --- and thus $\varepsilon_{min}$--- can be
about ten times larger.

However, no matter how low is this value, it is non-zero. This means that
if the quantity of available energy is limited, then the asymmetric
state will be able to be maintained only until energy is remaining,
that is during the interval time $\Delta t \leq E_{\rm tot}/\varepsilon_{\min}$.
Moreover, it is important to note
that the value of $\varepsilon$ is a \emph{minimum} prerequisite. It
represents only the energy that is dissipated by the racemization
process. Complete chemical systems are likely to require much more
energy, as each reaction of the full network, that is required for
the deracemization process, dissipates energy on their own. In the
example of the recycled system of activated
polymerization-depolymerization of amino acid
\cite{Plasson.Bersini.ea-04}, for which the energetic analysis has
recently been performed in detail \cite{Plasson.Bersini-09}, the
energy dissipated by the racemization reactions represents about only
$0.1\%$ of the total consumed energy.

\section*{Application to some example systems}
\label{sec:examples}

\subsection*{The Soai reaction: closed systems}
\label{sec:irreversible-systems}

The simpler way to generate a non-racemic system in the laboratory is
to work in a closed system, from an unstable initial state, and let it
evolve spontaneously toward a frozen state. The Soai reaction
\cite{Soai.Shibata.ea-95} describes exactly this process. The system
initially contains very reactive compounds, generating much more
stable compounds in a quasi-irreversible reaction. Due to a very
efficient autocatalytic mechanism, symmetry breaking can be observed,
forming an asymmetric alcohol with a very high enantiomeric excess,
starting from only symmetric compounds.

This final asymmetric state is a frozen state that only exists
temporarily. On larger timescales, taking into account the
reversibility of the reactions, there must be a long term racemization
of the system \cite{Crusats.Hochberg.ea-09}. By the time the system
has reached the asymmetric state, it has consumed all its energy, and
then it slowly goes back to the equilibrium state. The formation of an
asymmetric state is only possible as long as energy is available, that
drives the reactions in autocatalytic loops.

Reversible reactions have been invoked for describing the mechanism of
bifurcation in chemical systems based on the Mannich reaction
\cite{Mauksch.Tsogoeva.ea-07,Mauksch.Tsogoeva.ea-07*b,Mauksch.Tsogoeva-08},
presenting similar symmetry breaking behavior. This approach was
criticized, as reversible reactions must lead to the racemization of
the system in the long term
\cite{Blackmond.Matar-08,Blackmond-09*c}. However, if precautions are
taken into account, the presence of reversible reactions, as opposed
to irreversible ones, is not a major drawback for these systems
\cite{Plasson-08,Plasson.Bersini-09}. It is crucial to identify the
presence of energy: if energy is present in excess from the beginning,
then the reversible system can be able to reach a non-racemic state,
but this one will naturally racemize itself on a timescale that may
and may not be important; if energy is continuously exchanged with the
surroundings, then a non-racemic steady state can be maintained.

\subsection*{Generating asymmetry during crystallization: from closed
  to recycled systems}
\label{sec:cryst-syst}

Generating microscopic asymmetry from a symmetric system has been known
for a long time, using conglomerate crystallization of sodium
chlorate \cite{Kipping.Pope-98}. The same system was used later to
generate macroscopic asymmetry \cite{Kondepudi.Kaufman.ea-90}, the
bifurcation being generated by secondary nucleation induced by
stirring \cite{McBride.Carter-91}.

The problem with crystallization is that the initially ``bad seeds''
are not eliminated, as crystals are formed once and for all. More
recently, variations of this systems \footnote{There are systems that
  are either based on the original systems of sodium chlorate
  crystallization, or on a variant of amino acid crystallization in
  racemizing environment.} have allowed a total deracemization,
starting from a racemic mixture of homochiral crystals
\cite{Viedma-05,Viedma.Ortiz.ea-08}. In this system, some mechanisms are
present that can generate the destruction of crystals: a continuous
attrition of the crystals, due to a strong crushing by glass beads,
and a process of Ostwald ripening
\cite{Noorduin.Meekes.ea-08,Noorduin.Meekes.ea-08*b}.

\begin{figure}[htb]
    \centering
    \includegraphics[width=6cm]{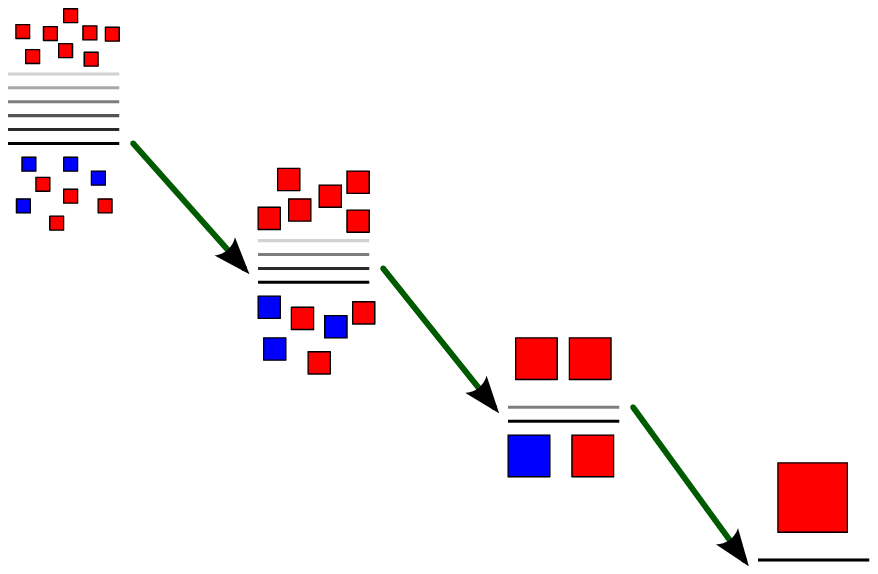}\\
    A: Ostwald ripening\\[1em]
    \includegraphics[width=6cm]{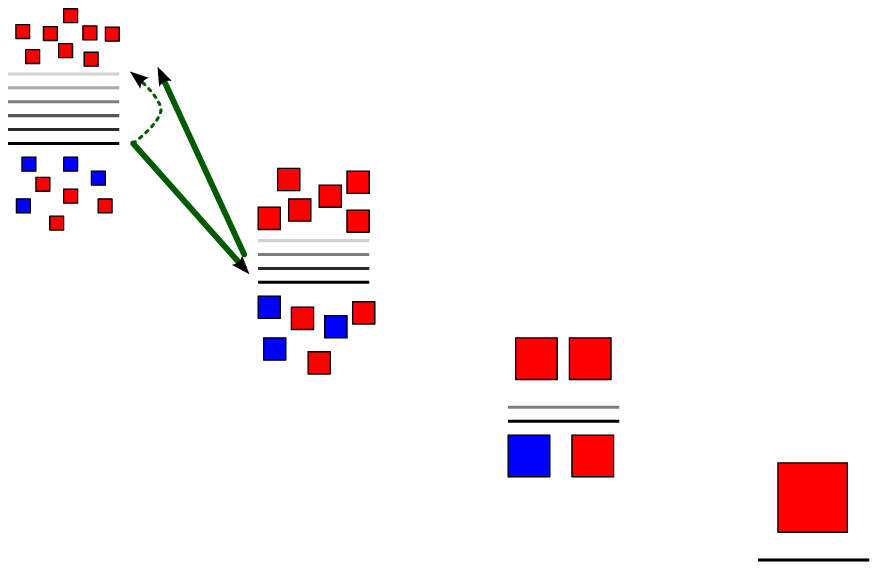}\\
    B: Grinded crystallization
  \caption{Energy profile for the evolution of a racemic mixture of
    conglomerate crystals by simple Ostwald ripening (A) and in
    grinded crystallization (B). A system containing a racemic mixture of homochiral
    crystals is entropically stabilized compared to an homochiral
    mixture of homochiral crystals, for the same size distribution in
    crystals of the same amount of matter. A system containing a low number of
    large crystals is enthalpically stabilized compared to a system
    containing a high number of small crystals, for the same amount
    of matter. Processes forming larger crystals are spontaneous,
    while processes forming smaller crystals require some consumption
    of energy.}
  \label{fig:cryst}
\end{figure}

The process of Ostwald ripening \cite{Voorhees-85} relies on the fact
that large crystals are more stable than small one. A system
containing several crystals of different sizes will evolve slowly, the
smaller crystals disappear while larger crystals are growing
\cite{Noorduin.Meekes.ea-08}. This is explained by realizing that,
eventually, one unique crystal will be obtained, and the system will
thus reach a frozen homochiral state, with one unique asymmetric
object that cannot evolve anymore (see Fig.~\ref{fig:cryst}.A).

In the Viedma experiments, the spontaneous Ostwald ripening process
(that tends to increase the size of the crystals) is counteracted by
additional processes that tend to decrease the size of the
crystals. As these processes generate less stable crystals, they must
be activated by processes consuming energy. The major process is a
system of continuous grinding, generated by the addition of crystal
beads \cite{Viedma-05}, implying a transfer of mechanical energy into
chemical energy. More recently, a similar effect was also induced by
temperature gradients \cite{Viedma.Ortiz.ea-08}.

In these systems, the Ostwald ripening is a downhill process (a
spontaneous evolution towards more stable state) where smaller
crystals are gradually destroyed in favor of the formation of larger
ones. This is an irreversible process, consuming the excess of
energy initially present in the system, characterized by an excess of
less stable small crystals. Simultaneously, the grinding process is an
uphill process, consuming the energy given by the mechanical grinding,
where the crystals are partially destroyed to generate smaller (and
thus less stable) crystals (see Fig.~\ref{fig:cryst}.B). The result is
the maintaining of a steady-state size distribution. This is not an
equilibrium state, as unstable compounds (small crystals) are
maintained by a continuous input of energy (the grinding). This
implies continuous cycles of reaction of crystal growth/decay,
leading to the deracemization of a stable set of crystals. This system
constitutes a recycled system as described in
Fig.~\ref{fig:stabl_non_rac}.C \cite{Plasson-08}.

It can be noted that the case of systems in the absence of attrition
crushing, and in the absence of temperature gradient, is reported to
be still working, but with a much lower efficiency. According to
Viedma \emph{et al}\cite{Viedma.Ortiz.ea-08}, the only possible processes of destruction
--- still present
as an homochiral state is obtained before a single object is formed
--- are probably collisions with other crystals, with vessel walls,
or with the stirring bar. Here, there is still a
system of growth by Ostwald ripening, while the mechanisms of crystal
breaking (and thus of energy consumption) are still present, but in
much lower quantity. No steady state size distributions are observed,
and the crystals are continuously growing. These systems are
intermediate cases between the two preceding ones, where part of the
energy comes from the Ostwald ripening process, and the other part
from the remaining crystal breaking mechanism.

\section*{Conclusion}
\label{sec:conclusion}

The emergence of a stable non-racemic state relies on the presence of a
usable energy source. Different energy consuming processes are
required:
\begin{description}
\item[First phase:] Creation of asymmetric objects. The asymmetric
  object can be a simple mole\-cule, but can also result from the
  association of other smaller objects (e.g. into polymers or
  crystal). This first step may not require energy consumption
  \emph{per se}, as an asymmetric object can be more stable than a
  symmetric object, but the created asymmetry is only local. Its
  extent is restricted to the dimension of the object itself. On
  larger scales, several objects can be independently created, leading
  to a racemic mixture of chiral objects. If a unique object is
  obtained, the system can be considered to be in thermodynamic
  equilibrium if the inversion of the object can be considered as
  impossible (e.g. in the case of one conglomerate monocrystal); when
  the racemization of the object is possible, the object is actually
  kinetically frozen.
\item[Second Phase:] Creation of a macroscopic asymmetric state. This
  requires either internal energy (the system starts from an initially
  unstable state, possessing an excess of chemical energy) or external
  energy (an external source of energy can be coupled to the system,
  forcing some reaction in a given direction). This energy must be
  directed toward some self-amplification mechanism, allowing the
  bifurcation toward one state over another. If the amount of energy
  is limited (especially in the case of a closed system functioning on
  an excess of internal energy), the created asymmetric state is
  unstable and is bound to be racemized with time.
\item[Third phase:] Stabilization of the asymmetric state. If the
  energy fluxes can be maintained, the deracemization processes can
  overcome the racemization processes and maintain actively a stable
  asymmetric state, even on a timescale larger than the racemization
  timescale. In this case, the deracemizing processes must consume at
  least the energy that is dissipated by the racemization process.
\item[Fourth phase:] Propagation of asymmetric regions in space: no
  additional energy is consumed to propagate a given asymmetry in
  space, on top of the energy consumption by each local systems, as
  long as simple diffusion can be sufficiently effective. However, the
  presence of active processes generating macroscopic turbulences
  inside the system can lead to a more efficient propagation. 
\end{description}
The kind of system described above is an obvious example of energy-driven
self-organizing system. In that respect, understanding the origin of
life implies to identify the potential energy sources, how this energy
can be used to generate bifurcation towards non-equilibrium states of
lower entropy, and how these states can be maintained. This energetic
approach is necessary to understand the evolution and stability of
self-organized chemical systems as they have been described until now
\cite{Eigen-71,Ganti-84,Farmer.Kauffman.ea-86,Wagner.Ashkenasy-09*b}.

\paragraph{Acknowledgment}
\label{sec:acknowledgment}
  We thank the European program COST ``System Chemistry''
  CM0703 for financial support. We also thank Nordita for the
  organization of the program ``Origins of Homochirality''. We
  especially thanks all the participants of this program and of
  the''Chirality'' working group of the COST program for the
  interesting discussions and debates about this subject.

\begin{appendix}
\section*{Appendix}

\subsection*{Energy transfer}
\label{sec:energy-transfer}

How chemical energy can be transferred into a system? The purpose is
to couple a thermodynamically unfavorable reaction
(e.g. $C\rightleftharpoons A$) to a more favorable one
(e.g. $X\rightleftharpoons Y$), so that the global reaction
is favorable (e.g. $C+X \rightleftharpoons A+Y$), forcing
the first reaction to be realized. Such transfer of chemical energy is
ubiquitous in biochemical systems, the most simple example being the
reaction $ATP \longrightarrow ADP + P_i$.

\begin{figure}[htb]
  \centering
    \includegraphics[width=5cm]{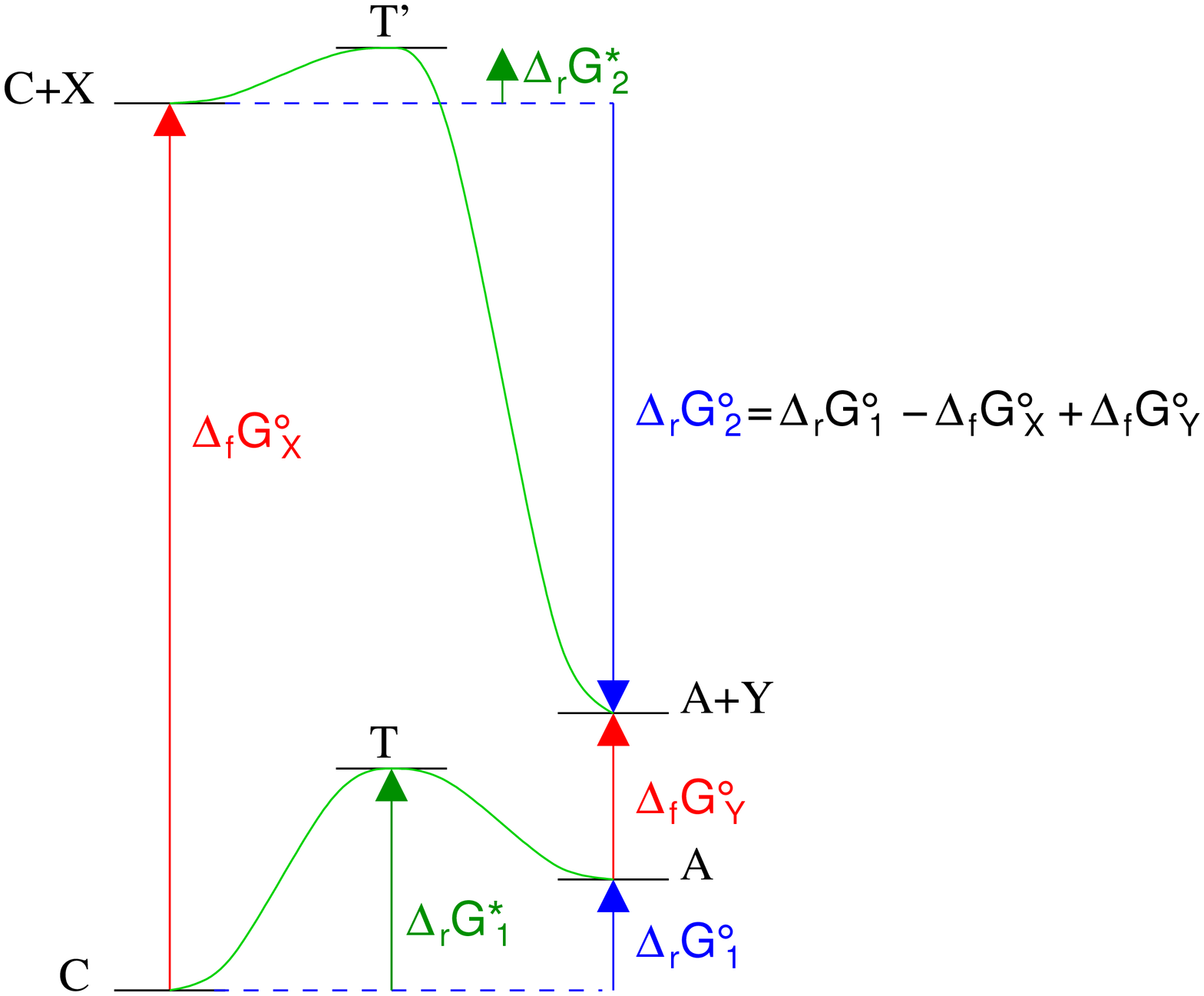}%
    \includegraphics[width=4cm]{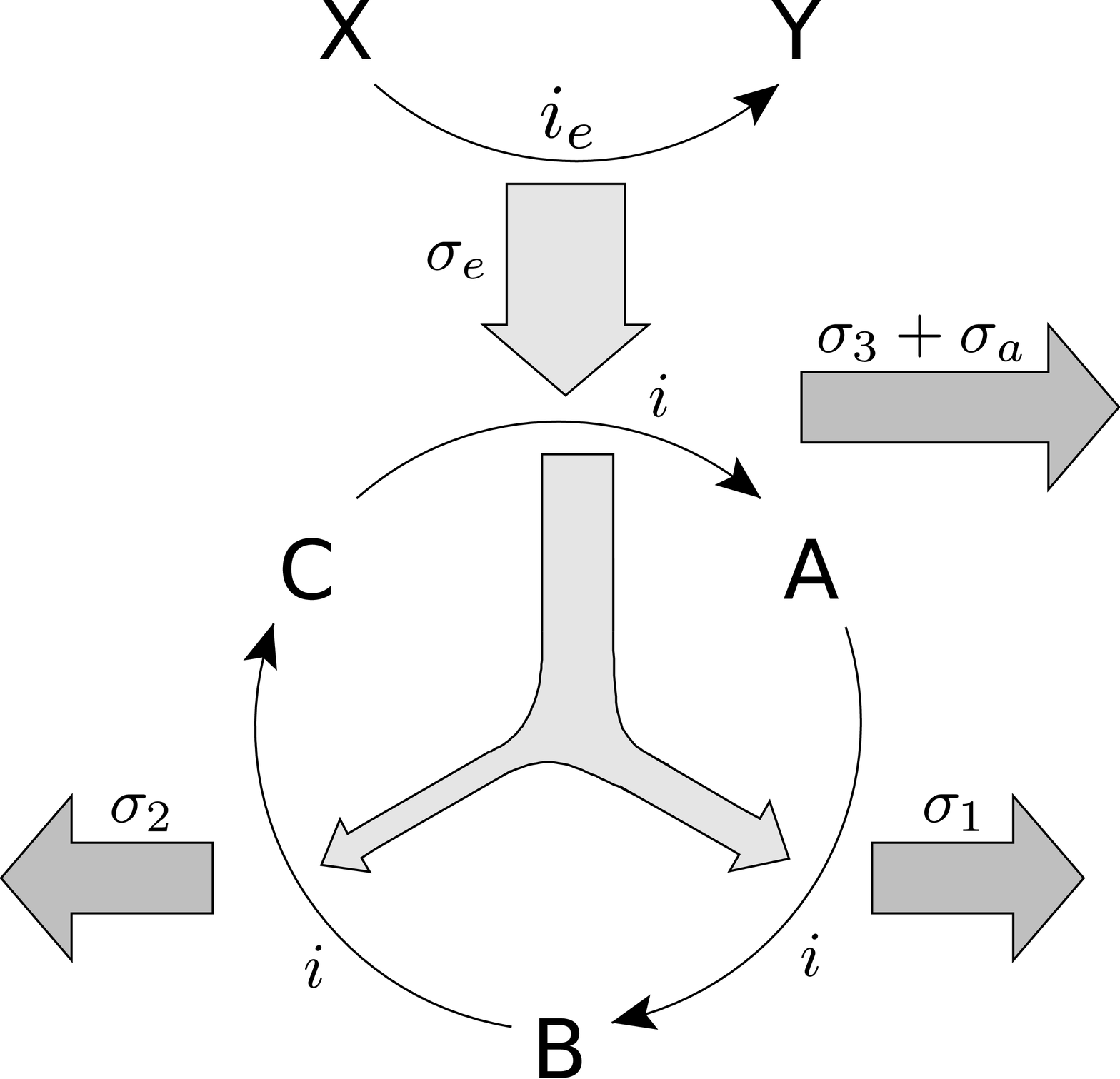}%
  \caption{How to transfer chemical energy. The thermodynamically
    unfavorable reaction $C \longrightarrow A$ is coupled to the more
    favorable reaction $X \longrightarrow Y$. The corresponding transfer
    of energy allows the generation of a reaction flux from $C$ to
    $A$, that can then react on its own generating larger cycle of
    reactions (e.g. a $A \longrightarrow B \longrightarrow C \longrightarrow A$ cycle.)}
  \label{fig:energ_transfer}
\end{figure}

If the system is connected to an energy reservoir fueling continuously
$X$ compounds, the spontaneous flux $X \longrightarrow Y$ will generate an
induced flux $C \longrightarrow A$, thus maintaining high concentrations of
the less stable compound $A$ (see Fig.~\ref{fig:energ_transfer}). Such
non-equilibrium state is characterized by internal chemical fluxes,
maintained by the coupling with external fluxes  \cite{Plasson.Bersini-09}.

\subsection*{Presence of free energy}
\label{sec:pres-energy}

The onset of non-equilibrium reaction network relies on the stable
presence of chemical energy. Real chemical systems are generally away
from the equilibrium state as they continuously communicate with their
surrounding. Diverse forms of available free energy can be found in
most abiotic environments, for example induced by the continuous
flux of light from the Sun, due to the chemical inhomogeneities of large
bodies (e.g. the possibility to obtain fluxes of reduced compounds
from the Earth crust  \cite{Huber.Wachtershauser-98,Cody-04}), or by
the presence of high temperature gradients at the surface of the early
Earth  \cite{Abramov.Mojzsis-09}. All these natural sources of energy
represent potential entry points of energy flux for the prebiotic
chemical systems.

\end{appendix}


\ifx\mcitethebibliography\mciteundefinedmacro
\PackageError{achemsoM.bst}{mciteplus.sty has not been loaded}
{This bibstyle requires the use of the mciteplus package.}\fi

\end{document}